\documentclass[sigconf]{acmart}

\usepackage{booktabs} 

\usepackage{algorithm}
\usepackage{algpseudocode}
\setcopyright{rightsretained}

\acmISBN{123-4567-24-567/08/06}

\acmConference[SIGIR 2019]{SIGIR 2019}{July 2019}{Paris, France}

\acmArticle{4}
\acmPrice{15.00}

\setcopyright{rightsretained}
\acmConference[SIGIR 2019 eCom]{ACM SIGIR Workshop on eCommerce}{July 2019}{Paris, France}
\acmYear{2019}
\copyrightyear{2019}
\makeatletter
\renewcommand\@formatdoi[1]{\ignorespaces}
\makeatother
\acmISBN{}

\begin{document}
\title{Ranking sentences from product description \& bullets for better search}

\author{Prateek Verma}
\affiliation{%
	\institution{Jet.com and Walmart Labs}
	\streetaddress{221 River Street}
	\city{Hoboken}
	\state{New Jersey}
	\postcode{07030}
}
\email{prateek.verma@jet.com}

\author{Aliasgar Kutiyanawala}
\affiliation{%
  \institution{Jet.com and Walmart Labs}
  \streetaddress{221 River Street}
  \city{Hoboken}
  \state{New Jersey}
  \postcode{07030}
}
\email{aliasgar@jet.com}

\author{Ke Shen}
\affiliation{%
  \institution{Jet.com and Walmart Labs}
  \streetaddress{221 River Street}
  \city{Hoboken}
  \state{New Jersey}
  \postcode{07030}
}
\email{ke.shen@jet.com}

\begin{abstract}
Products in an ecommerce catalog contain information-rich fields like description and bullets that can be useful to extract entities (attributes) using NER based systems. However, these fields are often verbose and contain lot of information that is not relevant from a search perspective. Treating each sentence within these fields equally can lead to poor full text match and introduce problems in extracting attributes to develop ontologies, semantic search etc. To address this issue, we describe two methods based on extractive summarization with reinforcement learning by leveraging information in product titles and search click through logs to rank sentences from bullets, description, etc. Finally, we compare the precision of these two models.
\end{abstract}

%
%


\maketitle

\section{Introduction}
\label{sec:introduction}

\begin{figure}
	\centering
	\includegraphics[width=8cm]{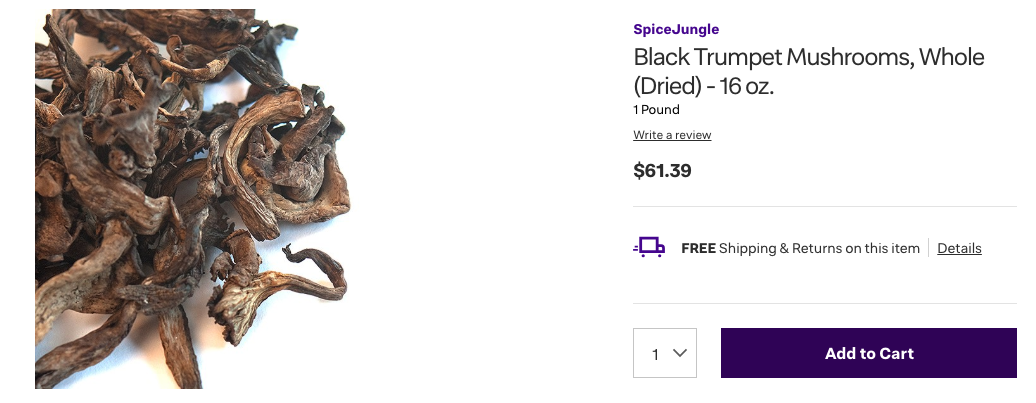}
	
	\centering
	\includegraphics[width=8cm]{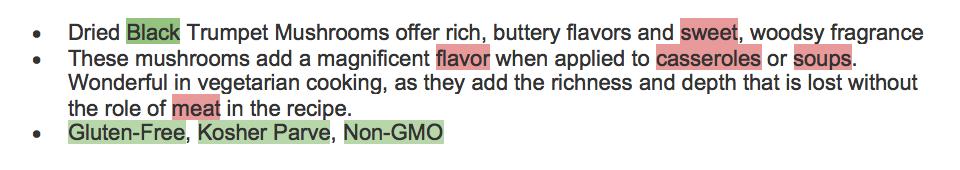}
	\caption{Sample SKU Image and Bullets}
	\label{fig:sku1Description}
\end{figure}

Many search engine frameworks like Solr ~\cite{solr} and ElasticSearch ~\cite{es} treat each sentence within a field in the document equally and this can lead to irrelevant documents present in the recall set. Consider Figure ~\ref{fig:sku1Description} which shows a sample item and some information associated in bullet form from an ecommerce website. The second bullet contains the terms ~\texttt{"soups"}, ~\texttt{"casseroles"} and \texttt{"meat"} because of which, the item (mushroom) will be present in the recall set for the search queries containing tokens like ~\texttt{"soups"} and ~\texttt{"casseroles"} due to full text match, leading to poor search relevancy. Relevant features for this SKU can be thought of as attributes that could be used in a search query to find this product. Thus, \texttt{"gluten free"} and \texttt{"non-GMO"} are considered relevant. Based on the attributes present in the sentence and how well it describes the item, we consider third bullet as more relevant to the item than the second bullet from search perspective. Tokens highlighted in Figure \ref{fig:sku1Description} with red and green color denote irrelevant and relevant features respectively for the SKU. \\

\begin{figure}
	\centering
	\includegraphics[width=9cm]{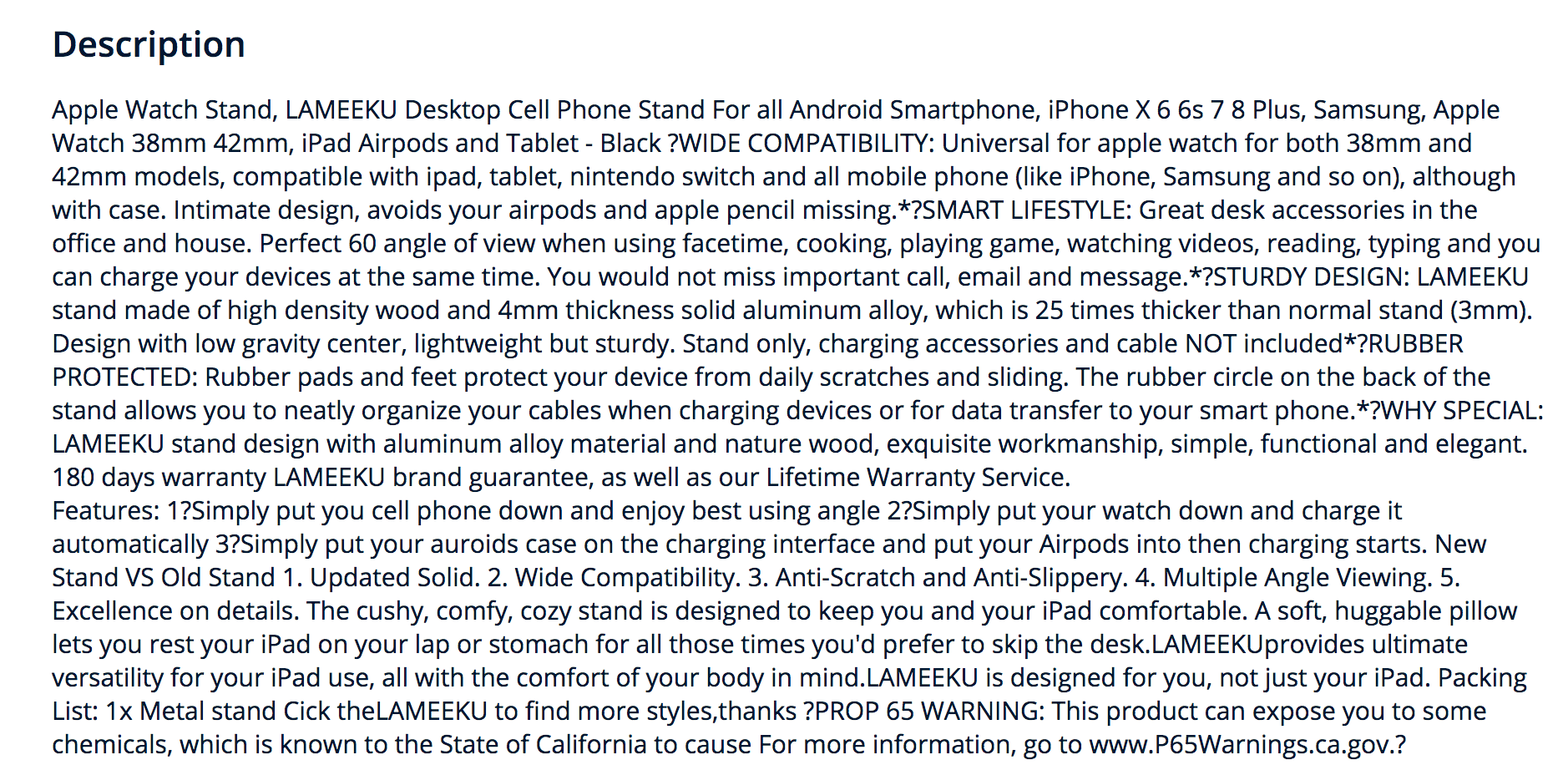}
	\caption{Sample SKU Description}
	\label{fig:sku2Image}
\end{figure}

\noindent Typically, this problem tends to appear in fields like product description and bullets which are often verbose and contain information about the SKU (stock keeping unit, a term used to describe item sold on the site) that is not pertinent to the item. Circuitous descriptions of the product and ~\emph{Keyword stuffing} are real concern in ecommerce. ~\emph{Keyword stuffing} refers to the practice of loading product data with keywords that may not be relevant with the item being sold. Figure ~\ref{fig:sku2Image}, which is a description of a SKU, illustrates this. \\

\noindent Product descriptions also tend to contain negations. That is, they describe what the product is NOT and what it is not suitable for. These kind of sentences are technically legitimate but poses a challenge for search engines and have the effect of returning misleading or irrelevant results. \\

\noindent A naive solution is to ignore these fields completely for search. While this may improve precision, it would be at the cost of recall, as relevant information might be lost. Such relevancy problems are mitigated by having semantic search using methods like query understanding. However, they require SKUs to have relevant attributes (atomic phrases that provide more information about an item ~\cite{DBLP:journals/corr/abs-1807-02039}) present in them to match it with user's intent. Thus, attribute extraction from the catalog data is often done in order to enrich SKUs (documents) with relevant attributes. \\

\noindent In this paper we describe a method to rank sentences based on if they are relevant from search perspective, and select top K sentences for search from these fields. Top K ranked set of sentences can lead to better full text match and can also help in extracting attributes for developing the ontology for semantic search ~\cite{DBLP:journals/corr/abs-1807-02039} as higher ranked sentences would have larger probability of attributes correctly describing the product. In our experiment, we limit K to 3. Thus, given a description of length greater than three, we always pick top three sentences generated by the model as our final summary. \\

\noindent Our contribution in this paper is, we demonstrate how ~\emph{Extractive Summarization} can be used to rank sentences present in product description and bullets using product title and user queries obtained from click through log. One of the benefits of this method is, cost of obtaining training data is cheap and the model can be run on items that have little or no click data associated with it. We also provide comparison of the two models by measuring precision@k of relevant sentences in the summary. \\

\section{Related Work}
\label{sec:relatedwork}

\noindent Summarization is the process of shortening a text document in order to create a summary while retaining major points of the original document. There are two kinds of summarization techniques: \emph{Abstractive} and \emph{Extractive summarization}. \emph{Abstractive summarization} involves using internal semantic representation and natural language generation techniques to create the summary ~\cite{chen2016distraction} ~\cite{rush2015neural}, ~\cite{see2017get}. \emph{Extractive summarization} involves selecting existing subset of words, phrases and sentences in the original text to generate the summary ~\cite{erkan2004lexrank}, ~\cite{nallapati2017summarunner}, ~\cite{yasunaga2017graph}. \\
 
\noindent Recently, a lot of work has been done on ~\emph{Abstractive Summarization} using attentional encoder-decoder model that was proposed by Sutskever et. al in ~\cite{sutskever2014sequence}. In  ~\cite{nallapati2016abstractive}, Nallapati et al. modeled abstractive summarization using Attentional Encoder Decoder Recurrent Neural Networks. While in ~\cite{paulus2017deep}, Paulus et. al introduced a new objective function that combined cross entropy loss with rewards from policy gradient reinforcement learning which improved state of the art in abstractive summarization.\\
 
\noindent~\emph{Extractive Summarization} was traditionally done using hand engineered features, such as sentence position, length ~\cite{radev2004mead}, words present in the sentence, their part of speech tags, frequency etc ~\cite{nenkova2006compositional}. However, with the recent success of encoder-decoder model, it is being used in ~\emph{Extractive Summarization} as well, such as ~\cite{cheng2016neural} ~\cite{narayan2017neural} and ~\cite{narayan2018ranking}. In ~\cite{cheng2016neural}, Cheng et al. developed a framework composed of hierarchical document encoder and attention based extractor for extractive summarization. In ~\cite{narayan2017neural}, Narayan et al. used the hierarchical encoder and attention based decoder to leverage side informations like title, image caption etc. and in ~\cite{narayan2018ranking} they introduced a new objective function based on ROUGE and used reinforcement learning to optimize it.\\

\noindent In this paper, we try to rank sentences using summarization techniques for the purpose of improving search relevancy. There hasn't been lot of work done in this area. One of the work that is aligned with our objective is from Ryen et. al ~\cite{white2002finding} published in 2002. They use statistical measures like frequency of query terms present in the sentence to rank them, and recommend user documents from the recall set by presenting them with ranked set of sentences for web search. However, our work focuses on ecommerce setting where we leverage Reinforcement Learning paradigm to rank sentences with the purpose of improving search by affecting recall/precision.

\section{Problem Formulation}
\label{sec:Problem Formulation}

\noindent Our objective is to rank sentences in product description and bullets from a search perspective. Search perspective means that when we extract attributes from sentences, they are relevant to the item and are likely to be used in a search query for that item. Methods like query understanding can benefit from ranked sentences as they use attributes in SKU to match with the user's intent. Higher ranked sentences are more likely to contain relevant attribute than a lower ranked sentences. Having a set of top ranked sentences would also help in full text match by avoiding queries to match with irrelevant sentences. We use \emph{Extractive Summarization} to achieve this. Our work is based on ~\cite{narayan2018ranking} which treats summarization task as a ranking problem and training is done by optimizing combination of ROUGE metric and cross entropy using reinforcement learning (described in \ref{ss:policylearning}). ROUGE stands for Recall-Oriented Understudy for Gisting Evaluation. It is a metric to compare automatically generated summary with the reference summary. ROUGE makes use of the count of overlapping units such as N-gram between the two summaries to measure the quality of system generated summary ~\cite{lin2004rouge}. Here we specifically use F1 score of three ROUGE scores mentioned below: \\
\begin{itemize}
	\item ROUGE-1: refers to the overlap of 1-gram between candidate summary and the reference summary (in our case title and queries)
	\item ROUGE-2: refers to the overlap of bi-gram
	\item ROUGE-L: measures Longest Common Subsequence based statistics to compute similarity between the two summaries\\
\end{itemize}

\noindent We use ROUGE because it is well aligned with our objective of finding relevant sentences from SKU description and bullets that is similar to the title and user engagement data (queries). It is the evaluation metric used in most summarization system, and training the model on a combination of ROUGE and cross entropy is shown to be superior than using just cross entropy ~\cite{narayan2018ranking}. REINFORCE algorithm is shown to improve sequence to sequence based text rewriting systems by optimizing non-differentiable objective function like ROUGE ~\cite{ranzato2015sequence} ~\cite{li2016deep}, so we use reinforcement learning to optimize our reward function.\\

\noindent We use title and queries obtained from click through log as part of the target summary. Title is one of the key fields in ecommerce catalog provided by the merchant, it captures essential information about the item and queries can be thought of as keywords users think are relevant attributes for the product. The intuition is, having them in the target summary would allow the model to capture important sentences present in the description and bullets. We create two models, one that uses just the title as target summary and the second model that uses top five queries that led to clicks on the item, along with the title as target summary. \\

\noindent Finally, we choose top K sentences as determined by the model as our final summary. Since, ecommerce product description tend to be short and less repetitive, the issue of repetition and diversity in not a concern in our summarization task.

\subsection{Network Architecture}
\label{sec:network architecture}

\begin{figure*}
	\centering
	\includegraphics[width=15cm]{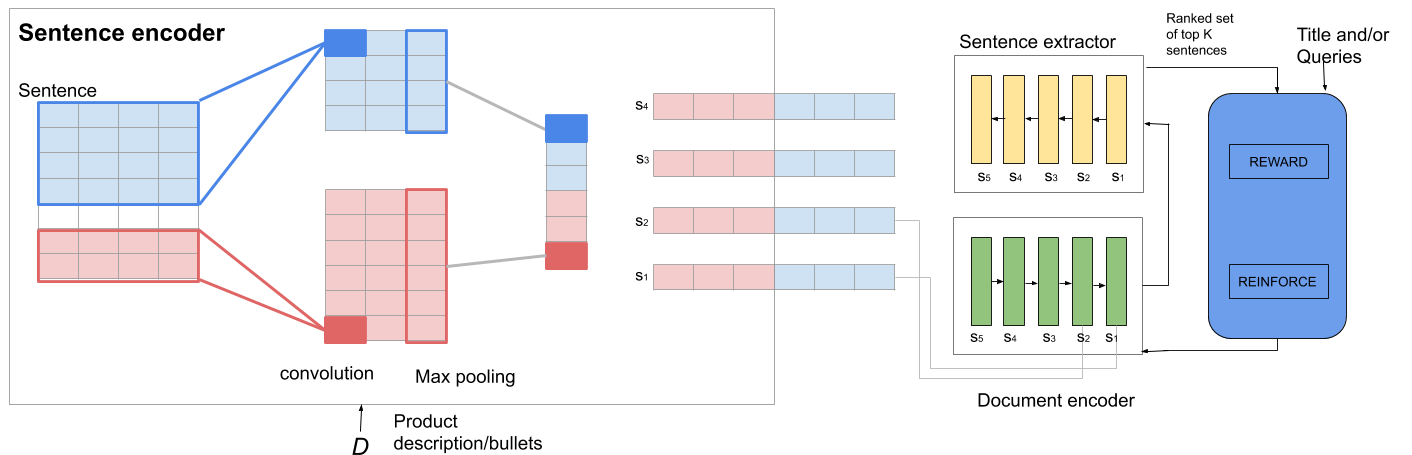}
	\caption{Network architecture}
	\label{fig:network_architecture}
\end{figure*}

Figure ~\ref{fig:network_architecture} depicts network architecture of the extractive summarizer. It aims to extract sentences \{$s_1..s_m$\} from a document $D$ composed of sentences \{$s_1..s_n$\} where $n>m$ and labels them 1 or 0 based on if they should be included in the summary or not. It learns to assign a score  $p(y_i|s_i,D,\theta)$ to each sentence which is directly proportional to its relevance within the summary. Here, $\theta$ denotes the model parameter, $s_i$ denotes the $i^{th}$ sentence and $D$ represents the document. Summary is chosen by selecting the sentences with top $p(y_i|s_i,D,\theta)$ score. Our network and the objective function is based on the paper ~\cite{narayan2018ranking}. We choose a sequence to sequence network which is composed of three main components: sentence encoder, document encoder and sentence extractor. \\

\noindent These components are described in detail below: \\

\noindent \textit{Sentence encoder} is composed of convolutional encoder which encodes a sentence into a continuous representation and is shown to capture salient features ~\cite{collobert2011natural}, ~\cite{kim2014convolutional}, ~\cite{kalchbrenner2014convolutional}. The encoding is performed using kernel filter $K$ of width $h$ over a window of $h$ words present in the sentence $s$. This is applied to each possible window of words in the sentence $s$ to produce a feature map $f \in R^{k-h+1}$, where $k$ is the length of the sentence. Then max pooling is performed over time on the feature maps and max value is taken corresponding to this particular filter $K$. Specifically, we use filter of size 2 and 4.\\

\noindent \textit{Document encoder}: The output of sentence encoder is fed to document encoder. It composes sequence of sentences to obtain a document representation. We use LSTM to achieve this. Given a document $D$ and sequence of sentence $(s_1…s_n)$ we feed sentences in reverse order to the model.
This approach allows the model to learn that the top sentences are more important and has been demonstrated in previous work ~\cite{sutskever2014sequence}, ~\cite{li2015hierarchical}, ~\cite{narayan2017neural}. \\

\noindent Finally, \textit{Sentence extracter} sequentially labels each sentence as 1 or 0 depending upon if the sentence is relevant or not. It is implemented using RNN with LSTM cells and a softmax layer. At time $t_i$, it makes a binary prediction conditioned on the document representation and previously labelled sentences. This lets it identify locally and globally important sentences. Sentences are then ranked by the score $p(y_i = 1|s_i,D,\theta)$. Here $s_i$ is $i^{th}$ sentence, $D$ is the document, $\theta$ is the model parameter and  $p(y_i = 1|s_i,D,\theta)$ is the probability of sentence $s_i$ being included in the summary. We learn to rank by training the network in a reinforcement learning framework optimizing ROUGE.
\\

\noindent We use a combination of maximum likelihood cross entropy loss and rewards from policy gradient reinforcement learning as objective function to globally optimize ROUGE. This lets the model optimize the evaluation metric directly and makes it better at discrimating sentences i.e it ranks the sentence higher if it appears often in the summary.

\subsection{Policy Learning} \label{ss:policylearning}

\noindent Reinforcement Learning is an area of machine learning where a software agent learns to take actions in an environment to maximize cumulative reward. It differs from supervised learning in the way that labelled input/output pairs need not be provided nor are sub-optimal actions need to be explicity corrected. Rather, the focus is on the balance between exploration and exploitation. 
Exploitation is the act of preferring an action that it has tried in the past and was found to be effective, whereas exploration is the act of discovering such actions, i.e. trying out actions that it has not selected before. \\

\noindent We conceptualize the summarization model in a reinforcement learning paradigm. The model can be thought of as an agent interacting with the environment, which consists of documents. The agent reads the document $D$ and assigns a score to each sentence $s_i \in D$ using the policy $p(y_i|s_i,D,\theta)$. We then rank and get the sampled sentences as the summary. The agent is then given a reward based on how close the generated summary is with the gold standard summary. We use F1 score of ROUGE-1, ROUGE-2, and ROUGE-L as the reward $r$. In our case, gold standard summary is the title and user queries. Agent is then updated based on the reward using the REINFORCE algorithm ~\cite{williams1992simple}. REINFORCE algorithm minimizes negative expected reward: \\

\begin{center}
	$L(\theta) = -\mathop{\mathbb{E}_{\hat{y}\sim p_\theta}}[r(\hat{y})]$ \\
\end{center}

\noindent Here, $p_{\theta}$ stands for $p(y|D,\theta)$, where $\theta$ is the model parameter, $D$ is the document and $r$ is the reward. \\

\noindent REINFORCE algorithm is based on the fact that the expected reward function of a non differentiable function can be computed as: 
\\

\begin{center}
	$\bigtriangledown L(\theta ) =  -\mathop{\mathbb{E}_{\hat{y}\sim p_\theta}}[r(\hat{y})\bigtriangledown {\log p}(\hat{y}|D,\theta)]$ \\
\end{center}

\noindent Calculating expected gradient in the above expression can be expensive as each document can have very large number of candidate summaries. It can be approximated by taking single sample $\hat{y}$ from $p_\theta$ for each training example in a batch, following which the above expression gets simplified to:
$$
\begin{matrix}
	\bigtriangledown L(\theta) &  \approx&  -r(\hat{y})\bigtriangledown \log p(\hat{y}|D,\theta)\\
	&\approx  &-r (\hat{y})\sum_{i=1}^{n} \bigtriangledown \log p(\hat{y_i}|s_i,D,\theta)
\end{matrix}	
$$

%

\noindent Since the REINFORCE algorithm starts with a random policy, and because our task can involve large number of candidate summaries for the document, training the model can be time consuming. So, we limit the search space $\hat{y}$ with smaller number of high probability samples $~\mathbb{\hat{Y}}$ consisting of top $k$ extracts. The way we choose these top $k$ extracts is, we select $p$ sentences which have highest ROUGE scores on its own and then generate all possible set of combination using these $p$ sentences with the constraint that maximum length of the extract can be $m$. We rank these against the gold summary using F1 score by taking mean of ROUGE-1, ROUGE-2 and ROUGE-L. We choose top $k$ of these ranked summaries as $~\mathbb{\hat{Y}}$. During training, we sample $\hat{y}$ from $~\mathbb{\hat{Y}}$ instead of $p(\hat{y}|\theta,D)$.

\begin{figure}
	\centering
	\includegraphics[width=9cm]{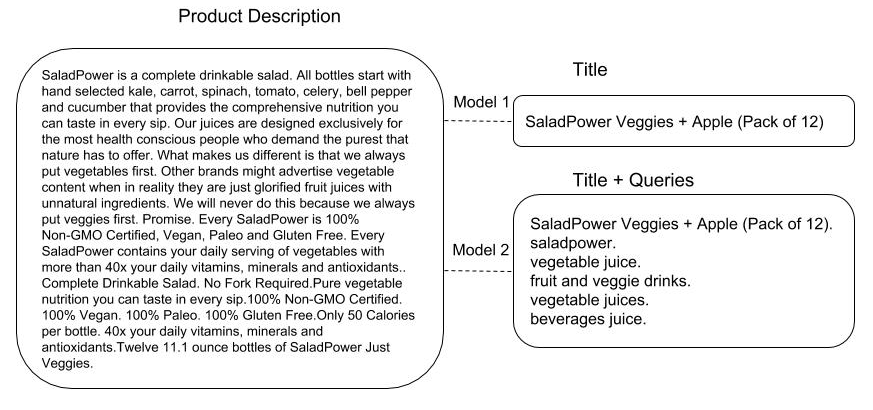}
	\caption{Retrieval by matching query understanding with SKU understanding}
	\label{fig:summary}
\end{figure}

\subsection{Input Data for model}
\label{sec:input_data}

\noindent We create two summarization models, one with title as its target summary (Model 1) and the other with title plus top five queries for which the product was clicked as the target summary (Model 2). Title and each query are treated as independent sentences when generating the reference summary. For input, we use product descriptions and bullets for both the models.\\
	
\noindent We preprocess the title, decription and queries before passing them to the model. Preprocessing step consists of sentence segmentation, tokenization, conversion of tokens into vocabulary id, truncation and padding to a fixed length. We use SKUs from grocery category of our catalog to evaluate the models. For ~\emph{Model 1} we used all the SKUs from the grocery category and for ~\emph{Model 2} we used a subset of SKUs from the category which had engagement above a certain threshold. Though ~\emph{Model 2} had fewer training data, it was richer since it had queries (top 5) associated with each SKU as part of the summary. One advantage of both methods is, it requires almost no manual effort to get the training data, thus is very cheap. Figure ~\ref{fig:summary} describes how the two models are set up for training.  \\

\noindent Since our objective is to have better full text match or attributes from the ranked set of sentences, 
each sentence can be independent of each other. This insight is well aligned with the framework of reinforcement learning based extractive summarization that optimizes ROUGE. \\

\begin{figure}
	\centering
	\includegraphics[width=8cm]{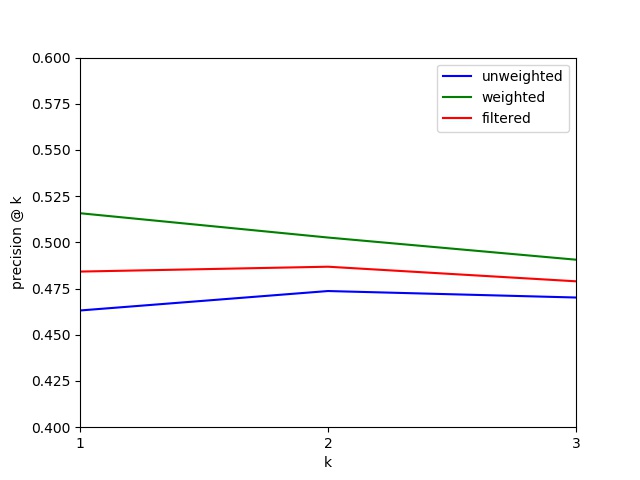}
	\caption{Precision @ k for the baseline}	
	\label{fig:baseline_precision}
\end{figure}

\section{Baseline Model}
\label{sec:baseline}

\emph{Tfidf} is one of the commonly used frequency driven approches for weighting terms to measure importance of a sentence for extractive summarization ~\cite{allahyari2017text}, ~\cite{neto2000document}. It measures the importance of words and identifies very common words in the documents by giving low weights to words appearing in most documents. The wieght of each word is computed by the formula: \\

\begin{center}
	$ \text{tfidf}(t,d,D) = \text{tf}(t,d) \cdot \text{idf}(t,D)$
\end{center}
\begin{center}
	$\text{idf}(t,D) = \dfrac{N}{|\{d\in D:t\in d\}|}$
\end{center}

\noindent Here, $\text{tf}(t,d)$ is the count of the term $t$ in the document $d$. \\
$\text{idf}(t,D)$ is the inverse document frequency. $N$ is the total number of documents in the corpus. ${|\{d\in D:t\in d\}|}$ is the number of documents where the term $t$ appears. If the term is not present in the corpus, it will lead to division by zero. To avoid this, it is a common practice to adjust the denominator to ${1+|\{d\in D:t\in d\}|}$.\\

\noindent For baseline, we use \emph{tfidf} based model. Our baseline consists of three aproaches that utilizes \emph{tfidf} to score the sentences to select top K. For the first approach, we sum up (unweighted) \emph{tfidf} score of the words to measure importance of a sentence and then select top K as the summary. Here, \emph{tf} is computed at the sentence level and \emph{idf} is across all the SKUs (documents).\\

\noindent For the second approach (weighted), we weigh the \emph{tfidf} score of tokens in the description that also appear in the title by multiplying it with a factor of $w_i$. The optimal wieght $w_i$ was found by using grid search method. In our case, it was found to be 2. \\

\noindent For the third approach (filtered), we sum up the \emph{tfidf} score of only those tokens in description that appear in the title. \\

\noindent Figure ~\ref{fig:baseline_precision} shows precision@k for the three models. As we can see from the graph, the weighted approach has highest precision@k, this shows that the words present in title does indicate which sentences are of relatively higher importance. However, it is also not a right strategy to exclude all the other words, as demonstrated by the higher precision@k of unweighted model over filtered model. Thus, in summary, boosting words present in title while also retaining other words for the computation of \emph{tfidf} score of a sentence seems to yield best result among all the baseline approaches.

\begin{figure}
	\centering
	\includegraphics[scale=0.25]{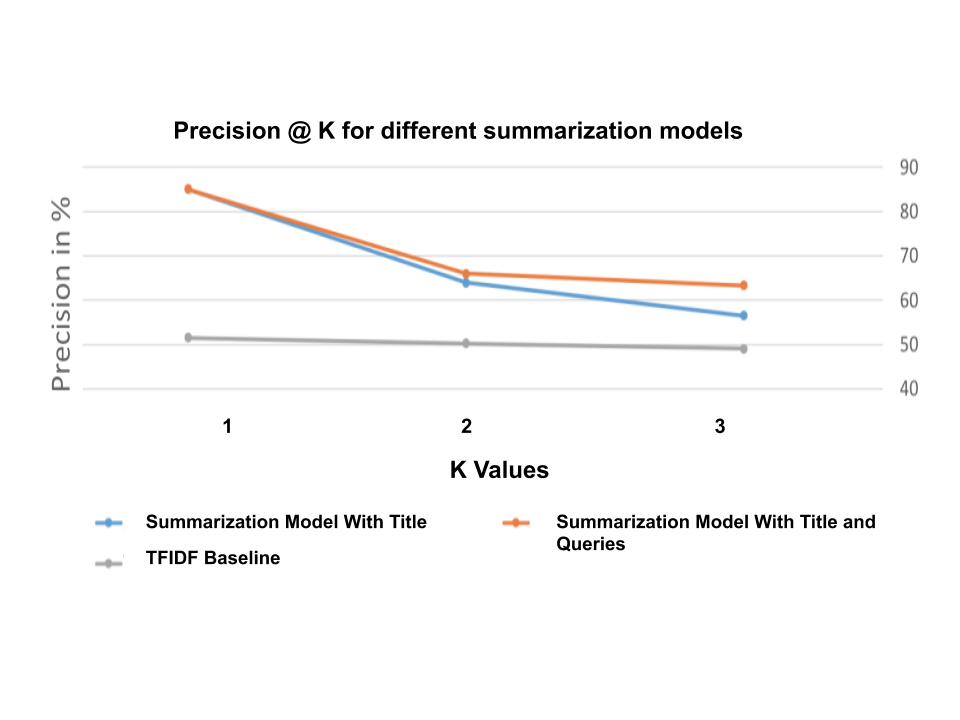}
	\caption{Precision @ k for Model 1(Title only), Model 2 (Title and queries) and the baseline}	
	\label{fig:precision_at_k}
\end{figure}

\section{Evaluation}
\label{sec:evaluation}

\noindent Our purpose of ranking is to find sentences that are relevant to the product and contain attributes of the product that customers might use in their search queries. This will improve results of full text match as well as query understanding, since it depends on matching user's intent with attributes extracted from the SKU. To analyze this, we reviewed 100 SKUs randomly sampled from the grocery category and manually labeled the sentences based on whether they were relevant or not. We evaluated the model using precision@k, with k as 1,2 and 3. \\

\noindent Based on the evaluation of the three \emph{tfidf} based models as described in the section \ref{sec:baseline}, we chose  \emph{weighted} Model (the second approach) as our baseline, as it has the best performance.\\

\noindent Figure ~\ref{fig:precision_at_k} shows precision@k for the two sequence to sequence based model and the baseline. Blue line indicates the model that was trained using just the title as target summary (Model 1), orange line indicates the model that was trained using title and top five queries that led to clicks on the SKU (Model 2) while, gray line is the precision@k for the baseline. We found that both Model 1 and Model 2 outperform the baseline.  Model 2 was better by 3.125\% and 12.08\% over Model 1 for precision@2 and precision@3 respectively. We believe the reason for Model 2 to outperform Model 1 is that queries provide additional context regarding which sentences are important and captures key information of the product, which is key to summarization.  \\

\noindent This demonstrates that words present in title capture key information of the product being sold. Title is provided by the merchant, so it provides merchant's point of view regarding what aspect of the product is important. Whereas, words present in user queries indicate the attributes of product that the user cares about. So combining these two sources of information is a good way to infer relevant sentences of description from a search perspective. Also, since not all SKUs (documents) have user clicks or may have comparatively less engagement data associated with it, creating a model leveraging title and click through log to find relevant sentences provides a way to generalize it to SKUs (documents) that have little or no engagement data. \\

\begin{figure}
	\centering
	\includegraphics[width=8cm]{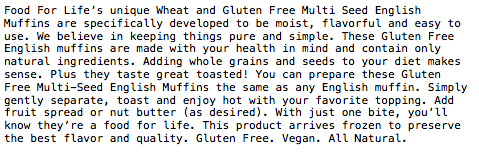}
	\caption{Input to the model: product description}	
	\label{fig:input_model}
\end{figure}

\begin{figure}
	\centering
	\includegraphics[width=8cm]{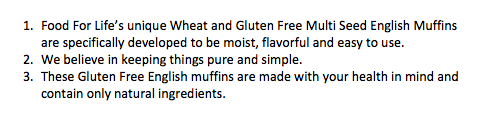}
	\caption{Model 1's output (title)}	
	\label{fig:model1_output}
\end{figure}

\begin{figure}
	\centering
	\includegraphics[width=8cm]{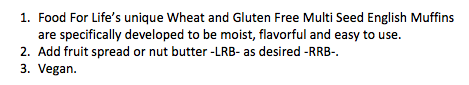}
	\caption{Model 2's output (title + query)}	
	\label{fig:model2_output}
\end{figure}

\noindent We provide one instance from our evaluation set as an example. Figure ~\ref{fig:input_model} shows a sample product description that is fed to the model. Figures ~\ref{fig:model1_output} and ~\ref{fig:model2_output} show output of Model 1 and Model 2 respectively.\\

\noindent Sentences that have keyword stuffing tend to be grammatically incorrect, structurally dissimilar to the title and generally longer. Thus, the intuition is that summarization models described above would rank such sentences lower. \\

\section{Conclusion and Future Work}
\label{sec:conclusion}

\noindent We implemented a framework to rank sentences from product description \& bullets based on ~\emph{Extractive Summarization} that uses reinforcement learning to optimize ROUGE and maximum likelihood cross entropy, thus enabling the model to learn rank the sentences. We compare two models, one that uses just the title and the other that uses queries from click through log along with the title. We show that these two models have higher precision in finding relevant sentences than the baseline which is a tf-idf based method to select top sentences. Typically, in search engines, such fields ~\texttt(product descriptions, bullets etc.) are either ignored or given a very low weight compared to fields like product title. Using this framework that ranks the sentences, we can assign a higher weight to ranked set of sentences. In addition, top N sentences from ranked set can also be used to extract attributes and help build the ontology. \\

\noindent Our future plan involves, 1) measuring the precision with two separate models, one for description and one for bullets, as they tend to have different grammatical structure 2) investigate the effect of query length on the ranking of sentences 3) have an algorithmic method to decide on the cut off (Top N) for selecting top sentences from each SKU. This is because, as length of the content in each SKU varies, number of relevant sentences could be different.

\bibliographystyle{ACM-Reference-Format}
\bibliography{sample-bibliography}
\end{document}